# Reversible Superdense Ordering of Tetragonal Lithium in a Layered Material

Natalie L. Williams[1,2 ‖], Stephen D. Funni[2 ‖], Sihun Lee[2], Drake Niedzielski[3], Mingjia Fang[2], Josh Leeman[4], Ratnadwip Singha[4,5], Saif Siddique[2], Tyler Hendee[6], Shiyu Xu[2,7], Jeffrey Kaaret[2], Giovanni Sartorello[8], Nicole A. Benedek[2], Leslie M. Schoop[4], Lynden A. Archer[9], Tomás A. Arias[3], Judy J. Cha[2,8*]

**Affiliations**

[1]Department of Chemistry & Chemical Biology, Cornell University, Ithaca, New York 14853, United States

[2]Department of Materials Science and Engineering, Cornell University, Ithaca, New York 14853, United States

[3]Department of Physics, Cornell University, Ithaca, New York 14853, United States

[4]Department of Chemistry, Princeton University, Princeton, New Jersey 08544, United States

[5]Department of Physics, Indian Institute of Technology Guwahati, Assam 781039, India

[6]Department of Engineering Physics, University of Wisconsin-Platteville, Platteville, Wisconsin 53818, United States

[7]Department of Mechanical Engineering and Materials Science, Yale University, New Haven, Connecticut 06511, United States

[8]Cornell Nanoscale Facility, Cornell University, Ithaca, New York 14850, United States

[9]Robert Frederick Smith School of Chemical and Biomolecular Engineering, Cornell University, Ithaca, NY 14853, United States

*Corresponding author. Email: jc476@cornell.edu

‖ These authors contributed equally.

**Abstract**

Understanding lithium (Li) ordering and dynamics is foundational in energy storage. X-ray based experimental methods do not simultaneously provide atomic structure information together with chemical composition and local bonding information for lithium in solids. Here we employ scanning transmission electron microscopy (STEM) and combine imaging, spectroscopy, and diffraction within a single experiment, to observe, *in situ*, an all-solid-state electrochemical cell. By integrating multimodal STEM with other complementary techniques, we report a complete mapping of lithium intercalation in a layered system, $LaTe_3$. We identify three ordered phases of $Li_xLaTe_3$ with x ranging from 1/3 to 3 with in-plane strain of up to 5%. At a very high lithium concentration of $Li_3LaTe_3$, we discover an unexpected three-layer, superdense lithium phase with tetragonal symmetry occupying the van der Waals gap. This represents a new Li phase that is reversible. Our multimodal approach thus enables complete tracking of lithium ordering and dynamics, important for next-generation energy storage applications.

## Introduction

Direct observation of lithium (Li) in solids during intercalation remains a longstanding experimental challenge. For *in-operando* battery studies, synchrotron characterization is widely used to track structural evolution during intercalation through methods such as X-ray diffraction (XRD) and X-ray absorption spectroscopy (XAS) (*1–5*). Probing lithium with X-rays, however, is difficult due to its low scattering power (*6*), inaccessibility via XAS, and limitations to incorporating multiple modalities in a single beamline experiment (e.g. transmission XRD and photoelectron spectroscopy). As a result, solving lithiated structures and quantifying lithium concentrations using X-ray based methods is usually indirect and requires prior knowledge and/or extensive modeling constraints (*7*). These constraints can complicate investigations of lithiated phases and likely underlie conflicting studies of lithium staging in graphite (*8*, *9*). A scanning transmission electron microscope (STEM) equipped with an energy loss spectrometer and a high dynamic range direct electron detector (*10*) uniquely overcomes this challenge. By offering imaging, diffraction, and spectroscopic modes in a single instrument for a single experiment, at up to atomic resolution, L.M. Brown's "synchrotron in a microscope" concept has been realized (*11*). Indeed, cryogenic TEM (*12–14*), liquid-cell TEM (*15–17*), and *in-situ* TEM (*18–20*) have been successfully employed for *in-situ* intercalation studies, sometimes combining multiple modalities. Here we adapt the design of Kühne et al. (*21*) to build a liquid-free electrochemical cell compatible with a standard biasing holder, and discover multiple, previously unknown intercalation phases within a single STEM experiment.

Utilizing *in-situ* multimodal STEM, we investigate lithium intercalation in $LaTe_3$, a layered system with an unconventional charge density wave (CDW) and square-lattice Te nets (*22*, *23*). From the combined datasets of real-space imaging, four-dimensional (4D)-STEM (spatially-resolved nanobeam electron diffraction), and electron energy loss spectroscopy (EELS), we observe three distinct lithium orderings in $Li_{1/3}LaTe_3$, $LiLaTe_3$, and $Li_3LaTe_3$ and map in-plane strains with values up to 5%. As EELS provides local bonding information in addition to composition, we reveal a gradual filling of the 5*p* orbitals in the square-lattice Te nets by lithium intercalation. *Ex-situ* electrochemical coin-cell experiments confirm the lithium concentrations and Raman measurements corroborate the distinct lithiated phases. Importantly, multimodal STEM reveals a superdense three-layer Li phase, which would be difficult to extract using X-ray-based methods. Density functional theory (DFT) calculations show that the lithiated structures are energetically stable and validate the superdense, tetragonal lithium phase in the host $LaTe_3$. This comprehensive, multimodal approach resolves coupled structural and electronic dynamics during intercalation in layered energy materials.

## Results & Discussion

We perform electrochemical lithiation on an exfoliated $LaTe_3$ flake using a Li anode and cured, polymer electrolyte inside a STEM (details in Methods). Fig. 1A shows a cross-section schematic of the *in-situ* electrochemical cell and Fig. 1B shows a top-down optical image of the cell prepared on an electrical TEM chip for *in-situ* STEM experiments. The polymer electrolyte only partially covers the flake to enable lithium intercalation while the uncovered portion remains electron transparent for *in-situ* STEM experiments. We used a $LaTe_3$ cathode as a model system for its unique square-lattice Te nets and quasi-two-dimensional (2D) structure (Fig. 1C). A high-angle annular dark-field (HAADF)-STEM image shows the pristine $LaTe_3$ flake (Fig. 1D) at open-circuit voltage (OCV) with the expected square lattice at atomic resolution (Fig. 1F). Prior to intercalation, the OCV was measured as 1.2 V. Applying −10 V to the cell produces changes in the HAADF intensity, suggesting successful intercalation (Fig. 1E). The system undergoes a complex structural and electronic evolution, including rearrangements of the 2D layer stacking, apparent by the altered lattice image shown in Fig. 1G. Fig. 1H–J shows systematic contrast changes in HAADF-STEM images of another $LaTe_3$ flake during lithium intercalation, indicating motion of a lithiation-induced phase front (Movie S1).

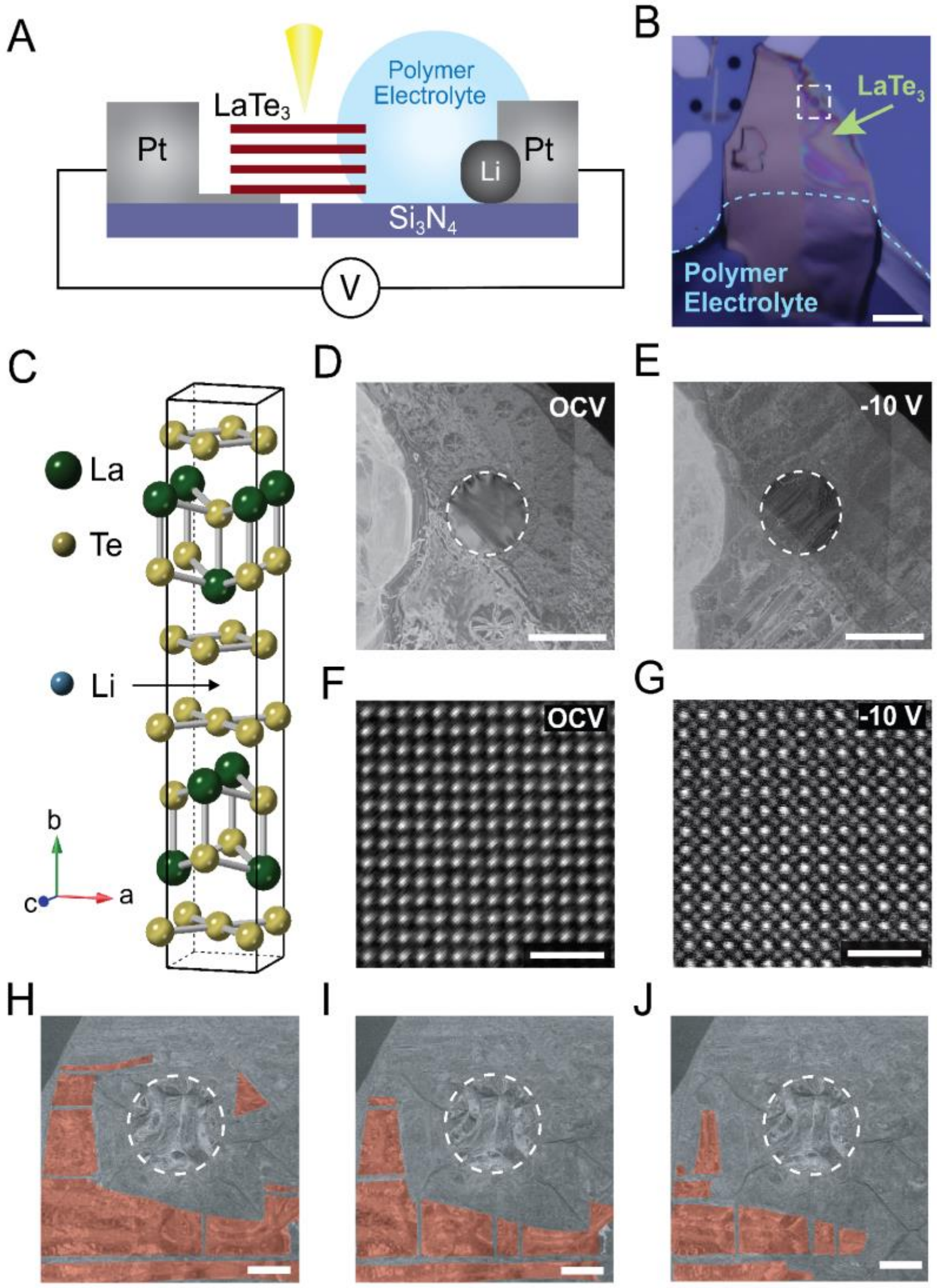

**Fig. 1. Lithium intercalation of $LaTe_3$.** (A) Cross-section schematic of a lithium-ion electrochemical cell (drawn not to scale). (B) Optical image of a $LaTe_3$ flake in an electrochemical cell partially covered by polymer electrolyte. Scale bar: 10 µm. White dashed square indicates the region of the flake shown in (D) and (E). (C) Structure of $LaTe_3$. Arrow denotes Li intercalation site. Low-magnification HAADF-STEM images of $LaTe_3$ at (D) OCV at 1.2 V and (E) −10 V. Scale bar: 5 µm. White dashed circle in (D) and (E) indicates the membrane hole. Atomic-resolution HAADF-STEM images at (F) OCV and (G) −10 V. Scale bar: 1 nm. (H–J) False-colored HAADF-STEM images taken 40 seconds apart of a different $LaTe_3$ flake during the early stages of intercalation. The colored region represents the initial un-intercalated region of the flake. White dashed circle indicates the membrane hole region under the flake. Scale bar: 2 µm. Full movie of progression of a lithiation-induced phase front, without false coloring, is shown in Movie S1.

4D-STEM datasets were acquired during lithiation, revealing distinct lithium orderings. In this technique, a ~ 4 nm-wide electron beam is scanned across the sample region, and electron diffraction patterns are acquired at each scan coordinate, yielding a 4D-dataset (*24*). Fig. 2A–D shows virtual dark-field STEM images reconstructed from the 4D-STEM data, and Fig. 2E–H shows the average electron diffraction patterns from the region of the flake over the $SiN_x$ membrane hole during lithiation. The pristine state at the OCV of 1.2 V displays the expected symmetry of $LaTe_3$, including the room temperature CDW that produces superlattice spots with the incommensurate ordering vector $q_{CDW} = {}^{2}/_{7}\mathbf{c}^*$ parallel to **c**-axis (*25*) (Fig. 2E). Upon lithiation by lowering the bias to 0 V to the flake, the CDW superlattice disappears and is replaced by a commensurate superlattice with q = ⅓**a*** or q = ⅓**c*** (Fig. 2F), indicating a new phase with lithium ordered inside the van der Waals (vdW) gap (Phase 1). Fig. S1 shows a line profile through the diffraction patterns indicating the different superlattice vectors. Here, the lithium ordering occurs primarily along the **a**-axis, but some domains along the **c**-axis are also observed, resulting in faint superlattice spots parallel to (002) in Fig. 2F. From the symmetry of the diffraction pattern, we propose a 1×1×3 supercell with composition $Li_{1/3}LaTe_3$ (Fig. 2N). A simulated diffraction pattern (Fig. 2J) from the proposed supercell matches the experimental data (Fig. 2F). The Li concentration is confirmed by EELS, which will be discussed later.

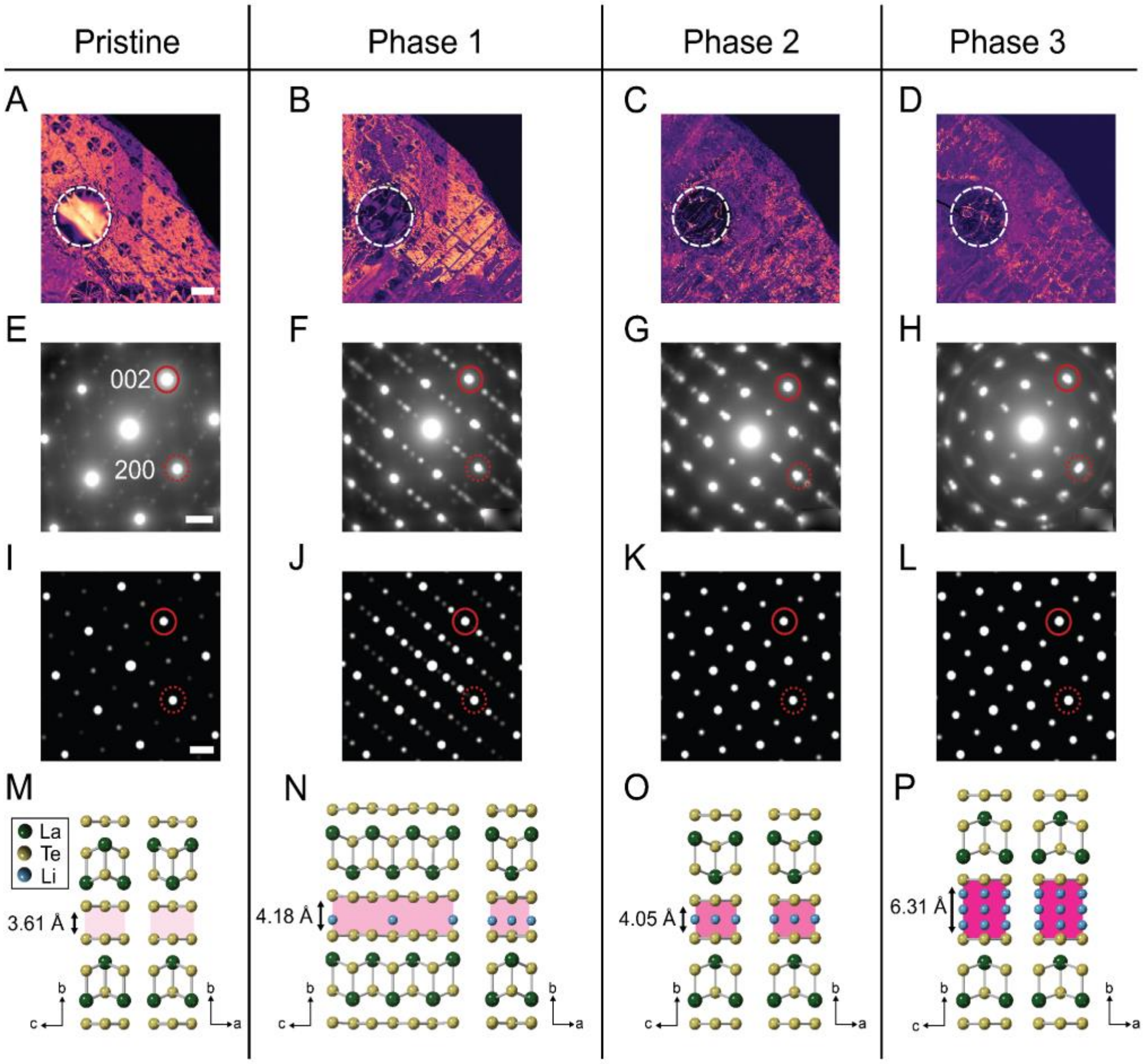

**Fig. 2. Lithiated phases of $LaTe_3$.** Virtual dark field images of the flake from the 4D-STEM datasets corresponding to the pristine (A) and Li-intercalated phases (B–D). Scale bar: 2 µm.

White dashed circle indicates the membrane hole under the flake. Average diffraction patterns (E–H) from the membrane hole region corresponding to (A–D). The central bright peak is the direct beam while the (002) and (200) peaks are indicated by solid and dotted circles, respectively. Simulated diffraction patterns (I–L) from the pristine (M) and DFT-relaxed structures (N–P), corresponding to each intercalated phase matching the experimental diffraction patterns. The Te nets are slightly corrugated in Phase 1. Scale bar: 0.2 $Å^{-1}$ for all diffraction images. Full 4D-STEM data during the lithiation is shown in Fig. S2.

Transition to Phase 2 occurs after ~200 min at the bias of −10 V, evident in the electron diffraction pattern (Fig. 2G) marked by the loss of the lithium ordering superlattice and increase in the intensity of kinematically forbidden reflections of the pristine structure (odd values of *h or l*). The latter indicates loss of symmetries with in-plane translational components (face-centering and glide plane) due to a shift of the interlayer stacking. A structure with the interlayer shift of ½ [001] and lithium concentration of $Li_{x=1}LaTe_3$ (Fig. 2O) produces a simulated diffraction pattern (Fig. 2K) in agreement with the experiment (Fig. 2G). We note that only a structure with shifts in interlayer stacking replicates the observed diffraction pattern; keeping the original stacking while inserting lithium in the vdW gap does not reproduce the experimental observation. The interlayer shift allows the positively charged lithium ions to occupy square-prismatic interstitial sites between the two Te square-nets.

Phase 3 emerges upon further intercalation after lowering the bias to −30 V. Phase 3 retains the pseudo-tetragonal symmetry of Phase 2, first appearing at a lithium concentration of $x \approx 2$ and showing stability to at least $x \approx 3$ (as quantified by EELS in Fig. 3). Fig. 2P shows the proposed structure of Phase 3, with a three-layer Li in the vdW gap. The three-layer Li is arrived at by analyzing the fine structures of EELS Li, La and Te edges along with 4D-STEM and high resolution imaging data, which is subsequently discussed in detail. This structure produces a simulated diffraction pattern (Fig. 2L) in agreement with the experiment. The full 4D-STEM data and atomic-resolution HAADF-STEM images are shown in Fig. S2–S4 and the lithiation experiment was reproduced for another $LaTe_3$ flake up to Phase 2 (Fig. S5).

DFT calculations were carried out at different lithium concentrations to support the experimental observations (calculation details can be found in Methods). The supercell structures shown in Fig. 2N–P are fully relaxed, energetically stable structures by the DFT calculations. The relaxed structure of Phase 1 shows corrugated Te square-nets, and the superdense three-layer Li in Phase 3 is reproduced in calculations. In the DFT model, this new Li phase is tetragonal with two equal in-plane axes and a slightly smaller out-of-plane axis compared to the body centered cubic Li metal (*26*). Calculations suggest additional lithiated structures are possible for Phases 1–3, and Fig. S6 shows simulated electron diffraction patterns for the possible structures of Phase 1. Complete DFT results of the favorable lithium sites (Fig. S7, S8) and additional structures (Fig. S9) with their corresponding electronic band structures (Fig. S10), phonon dispersions (Fig.

S11), and projected density of states (DOS) (Fig. S12) are shown in the Supplementary document, along with their tabulated formation energies (Tables S1–S3).

The unique advantage of STEM is the simultaneous acquisition of compositional and structure information within a single experiment. Fig. 3 summarizes the chemical evolution during lithiation obtained from EELS spectrum images, taken in between 4D-STEM data acquisitions. Fig. 3A–C show the *in-situ* evolution of core loss edges for Li, Te, and La, respectively, during lithium intercalation. Each spectrum is integrated from the membrane hole region and plotted in black, purple, red, and green to denote the pristine, Phase 1, 2, and 3 states identified from the 4D-STEM data.

The Li-K edge grows with more lithiation as expected (Fig. 3A). We quantify the lithium stoichiometry ($Li_xLaTe_3$) by comparing the Li-K edge to the La-$M_{4,5}$ edge. In Phase 1 (purple spectra), a slight lithium edge at 57 eV is observed, which sits in the extended fine structure region of the Te-$N_{4,5}$ edge. Elemental quantification (details in Methods and Fig. S13) indicates $x < ⅓$. In Phase 2 (red spectra), the Li-K edge increases significantly in intensity and develops a well-formed near-edge structure with a sharp peak below 60 eV. The fine structure is similar to that seen for LiF, indicating highly ionic bonding (*27*). We simultaneously observe a decreased intensity of the Te-$N_{4,5}$ edge onset. $N_{4,5}$ excitations originate from the 4*d* core shell with transitions to the empty 5*p* shell allowed by the dipole selection rule. Thus, we interpret the decreased intensity as filling of previously unoccupied Te 5*p* states by electron doping from Li, in line with the Te *p*-orbital character of the bands near the Fermi level in $LaTe_3$ (*25*, *28*). There is no actual loss of Te as the Te-$M_{4,5}$ edge remains unchanged (Fig. 3B). Additional EELS data for the reproduced lithiation experiment (Fig. S5) is shown in Fig. S14.

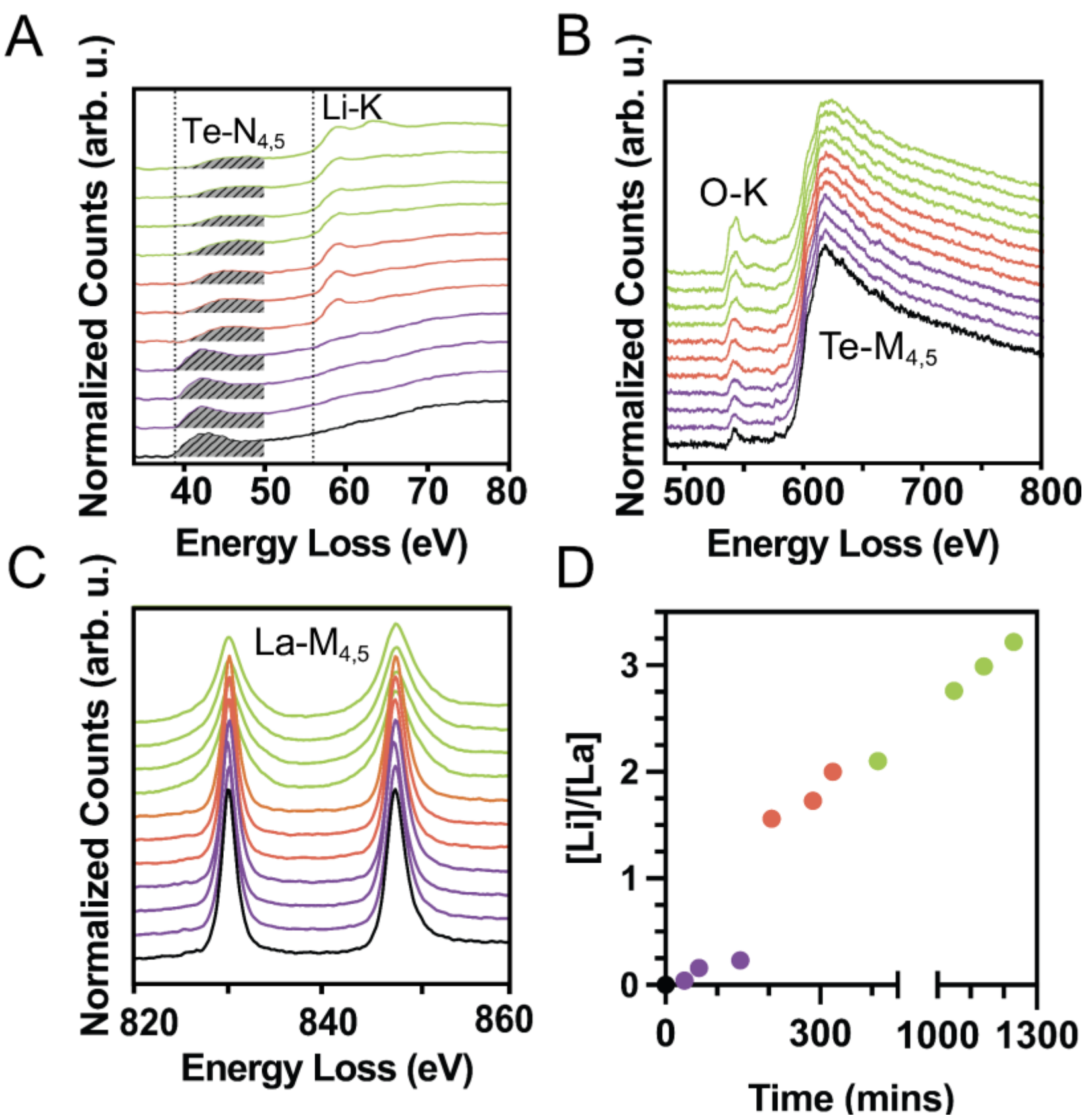

**Fig. 3. *In-situ* STEM-EELS measurements throughout lithiation.** (A) Low-energy core loss range containing the Te-$N_{4,5}$ and Li-K edges. (B) O-K and Te-$M_{4,5}$ edges. Note the decrease in intensity of the Te-$N_{4,5}$ edge onset (40 eV) with increasing Li concentration (A) despite no loss of Te as indicated in (B). (C) La-$M_{4,5}$ edge shows no chemical shifts or branching ratio changes indicating the La bonding environment is unchanged even at high lithium content. (D) Li concentrations relative to La as quantified from the spectra in (A) and (C). Data are color-coded to indicate different lithiated phases.

Transition to Phase 3 occurs at $x \approx 2$ (green spectra). It is characterized by a less prominent Li-K near-edge peak compared to Phase 2. Decreased ionicity in favor of more metallic Li-Li bonds would lead to this (*27*), suggesting a changed lithium bonding environment from Phase 2. The Te-$N_{4,5}$ edge onset is further suppressed due to increased filling of the Te 5*p* orbitals. While Li-K edge and Te-$N_{4,5}$ edge evolve during lithiation, Te-$M_{4,5}$ edge and La-$M_{4,5}$ edge do not change for Phase 3 (Fig. 3B, C). Fig. 3D plots the [Li]/[La] atomic ratio throughout the intercalation. Elemental maps for Li, Te, and La (Fig. S15) show homogeneous distribution of lithium.

We return to the discussion of the atomic structure of Phase 3. For Phase 1 and 2, 4D-STEM data clearly show distinct structures (Fig. 2F,G). In contrast, the electron diffraction pattern of Phase 3 (Fig. 2H) appears similar to that of Phase 2. To discern the atomic structure for Phase 3, several simultaneous observations must be interpreted self-consistently. The primary experimental evidence for Phase 3 is changes to the Li-K and Te-$N_{4,5}$ energy loss near edge structure (ELNES) that occur at x $\approx$ 2. ELNES can be interpreted as representing the local density of unoccupied electronic states, thus it is sensitive to changes in the bonding environment of the element probed. For Phase 3, the Li and Te edges exhibit ELNES changes (Fig. 3A), but La does not (Fig. 3C). This means that bonding environment is changing for Te atoms in the Te square-nets, reflecting lithium insertion in the vdW gap, while the unchanged La ELNES means bonding around the La-Te block remains unperturbed and indicates no lithium insertion into the La-Te block, consistent with DFT calculations (Fig. S9, Table S3). At the same time, electron diffraction (Fig. 2G–H, Fig. S2) and atomic resolution HAADF imaging (Fig. 1G, Fig. S4) are consistent with an altered layer stacking but offer no evidence of other significant structural rearrangement. Taken together with the EELS stoichiometry of x up to 3 (Fig. 3D, Fig. S13), these observations reveal a new superdense phase having multiple layers of lithium contained within the vdW gap. Fig. 2P shows the DFT-relaxed structure of Phase 3, with a three-layer Li with tetragonal symmetry in the vdW gap. This structure produces a simulated diffraction pattern (Fig. 2L) in agreement with the experiment.

The three distinct lithium orderings of $Li_{1/3}LaTe_3$, $LiLaTe_3$, and $Li_3LaTe_3$ identified from multimodal *in-situ* STEM are corroborated by *in-situ* Raman spectroscopy. For Raman measurements, the polymer electrolyte fully covers the flake. Optical images show the emergence of dark lines in the flake as a function of electrochemical potential, $V_{EC}$ (V) vs. $Li^+$/Li (Fig. 4A). These lines represent wrinkles in the flake, reflecting a mechanical response to strain generated during lithium insertion (*29*, *30*) (Fig. S16). A Raman spectrum from the pristine state shows four phonon modes at ~73, 88, 97, and 108 $cm^{-1}$ (Fig. 4B, black spectrum), consistent with previous reports (*22*, *31*–*33*). The modes at 73 and 88 $cm^{-1}$ couple strongly to the CDW (*34*). In Phase 1 at $V_{EC}$ = 1.0 V, these two modes disappear, suggesting CDW collapse, and new Raman modes appear at ~79, 86 and 92 $cm^{-1}$ indicating reduced lattice symmetry via Brillouin-zone folding from the lithium ordering superlattice (*35*). This agrees with the structure analysis from 4D-STEM for Phase 1, which showed the disappearance of CDW superlattice spots and an increase in unit cell size. Concurrently, additional Raman peaks appear at 120 and 142 $cm^{-1}$, resembling the $A_1$ and $E_2$ modes of elemental Te (*36*, *37*), likely reflecting distortions of the Te square-nets that cause inactive modes to become Raman-active (*38*). This is in line with perturbation of Te nets found by DFT calculations (Fig. 2N).

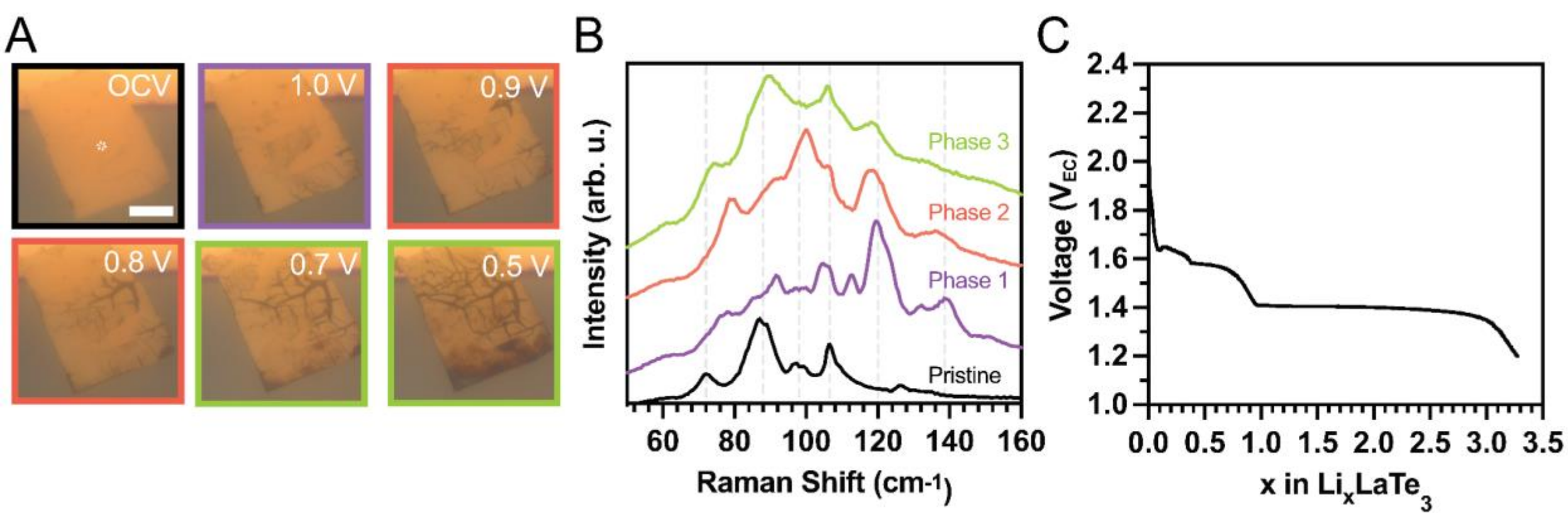

**Fig. 4.** ***In-situ*** **Raman characterization and galvanostatic discharge of $LaTe_3$ during lithiation.** (A) Optical images of a $LaTe_3$ flake under polymer electrolyte. The white circle marks the Raman collection spot for all spectra. Colored borders correspond to select spectra in (B). Scale bar: 10 µm. (B) *In-situ* Raman spectra of $LaTe_3$ with distinct lithiated phases shown in purple, red, and green. (C) Galvanostatic discharging of a $LaTe_3$ powder coin cell at a current of 0.071 mA and a current density of 0.1 mA/cm$^2$ in the voltage window 2.2 to 1.2 V vs. $Li^+$/Li.

In Phase 2 at $V_{EC}$ = 0.9 V, the two Raman modes at 120 and 142 cm$^{-1}$ red-shift and phonon modes between ~85 and 110 cm$^{-1}$ coalesce into a broad band around 100 cm$^{-1}$, suggesting increased structural heterogeneity (*39*). In Phase 3 at $V_{EC}$ = 0.7 V, the Raman spectrum partially recovers phonon frequencies of the pristine lattice but with broadened linewidths, pointing to local bonding environments that converge toward the original structure with growing disorder (*40*, *41*). This agrees with the electron diffraction of Phase 3, where the superlattice spots associated with lithium order disappear (Fig. 2H). Further lowering $V_{EC}$ leads to more peak broadening and intensity loss, indicating breakdown of long-range crystalline order (Fig. S17). The corresponding $V_{EC}$ vs. time plot for Fig. 4B is provided in Fig. S18. Thus, Raman measurements confirm the three distinct lithium ordering phases, in agreement with the STEM results. We note that the $V_{EC}$'s for phase changes are different between the Raman and STEM experiments because of the different configuration and amount of the polymer electrolyte used for intercalation. *In-situ* Raman experiments were repeated on multiple flakes (Figs. S17, S19–21). Scanning electron microscopy (SEM) (Fig. S16) and focused ion beam (FIB) (Fig. S22) cross-sectional imaging further support the inference of structural disorder during lithiation.

Galvanostatic discharge measurements of a $LaTe_3$ powder coin cell (Fig. 4C) also show three distinct and reversible lithiated states with phase boundaries at x = ~0.3, 1, and 3 Li per $LaTe_3$. This agrees with the EELS quantification of Li concentrations shown in Fig. 3D. A 4th phase with reduced reversibility is also observed in cycling and likely corresponds to the amorphous phase. Additional electrochemical measurements demonstrate stable cycling and capacity, consistent redox features, and reversible lithium transport at multiple scan rates and current densities (Figs. S23–26), confirming the robust phase evolution and reversibility of lithium-ordered phases.

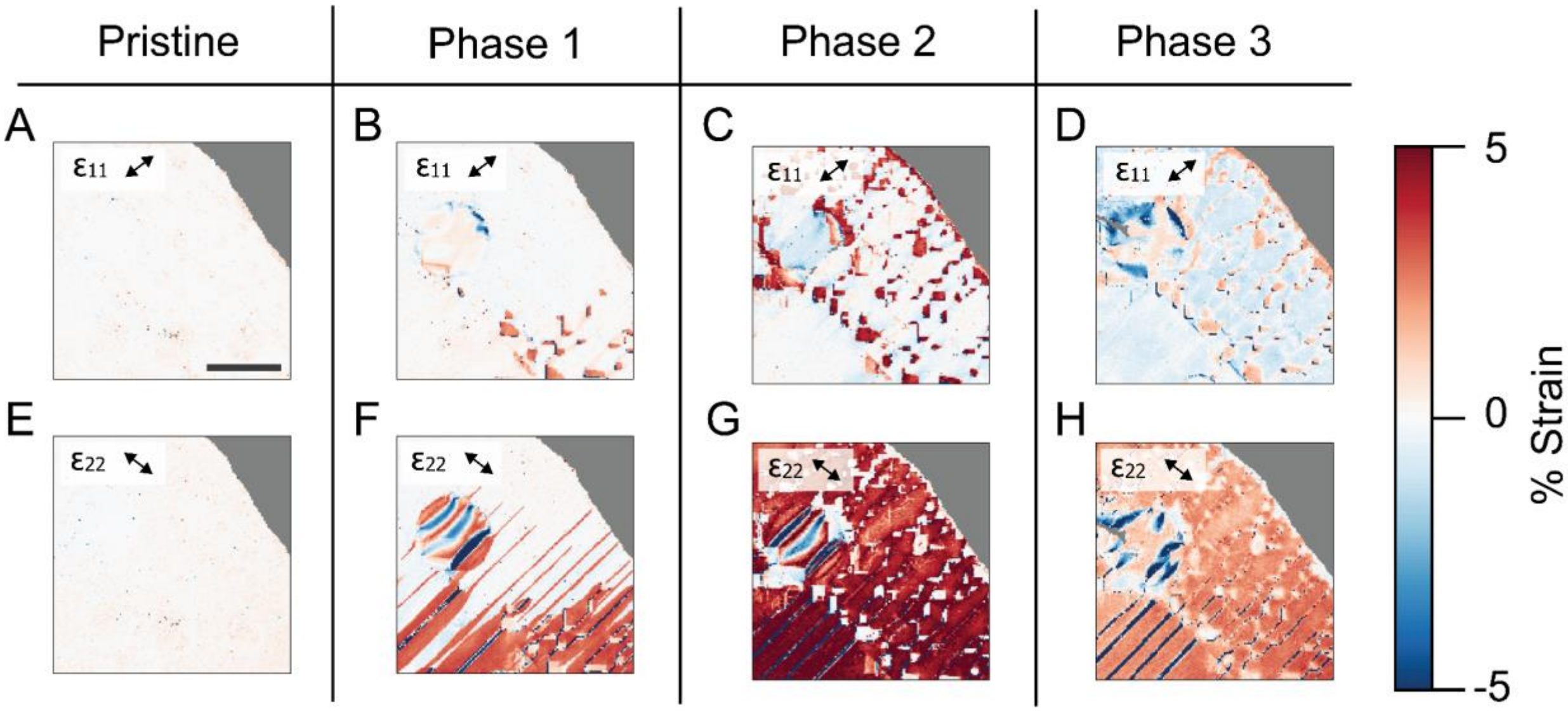

**Fig. 5. Strain in pristine and intercalated $LaTe_3$.** Representative strain maps along the **c**-direction (A–D) and **a**-direction (E–H). The average lattice from the membrane hole region of the pristine flake is used as the zero-strain reference for all maps. Note that: in (B, F) the lower portion of the flake and membrane hole areas have transitioned to Phase 1 while the remainder of the flake is in the pristine state; in (C, G) the entire flake has been lithiated with the lower portion in Phase 2 and the upper portion in Phase 1. Areas with large compressive strain values (dark blue) are attributable to the flake being locally off zone axis due to buckling folds induced from intercalation stresses. Scale bar: 4 µm. Movie S2 shows the full evolution of the strain change with intercalation.

Intercalation produces volumetric expansion in the host material. From our 4D-STEM data (Fig. 2), we measured the in-plane strains of the intercalated phases, finding that they are generally large and tensile (Fig. 5). Phase 1 shows unidirectional strain of approximately 2% along the local lithium ordering direction, with near-zero strain perpendicular to the ordering, such that the domains are apparent from the strain maps (Fig. 5B,F). The strains in Phase 1 agree well with the DFT result (Fig. 2N), which yields 0.96% strain along the ordering direction and no strain perpendicular. In Phase 2, when the lithium ordering is lost, the prior domains remain imprinted on the strain state (Fig. 5C,G), albeit with an increased unidirectional strain of around 5%. The DFT result at $x = 1.0$ (Fig. 2O) yielded 3.4% biaxial strain, roughly matching the experimental magnitude for Phase 2.

Surprisingly, Phase 3 shows a significant and global *decrease* in strain to ~1% along the previous ordering direction while the orthogonal component becomes slightly compressive (Fig. 5D,H). This may be due to changes of the $LaTe_3$ layer stacking or to altered bonding between the Te nets and three-layer lithium as the latter becomes increasingly metallic in the superdense phase. Immediately following the strain relaxation seen in 4D-STEM, atomic resolution HAADF-STEM imaging (Fig. 1G) shows the flake in an intermediate stacking arrangement, indicating some layers are shifted by ¼ per unit cell. This clearly shows that $LaTe_3$ undergoes layer stacking changes during intercalation. At the end of the experiment, however, HAADF-STEM images show that a regular stacking is recovered (i.e. layers shifted by either 0 or ½ unit cell along **a** or **c**). Fig. S2 and Movie S2 show the complete strain evolution throughout the experiment measured from all acquired 4D-STEM datasets. The out-of-plane strains could not be accurately measured experimentally, due to the layer stacking shifts that complicated the analysis of high-order Laue Zone in electron diffraction patterns (Fig. S27). DFT calculations find a significant opening of the vdW gap across the $Li_xLaTe_3$ lithiation sequence (Fig. 2M–P). The DFT calculation results of the stacking arrangement at various phases (Table S1–S3) are in accordance with our experimental observations.

**Conclusion**

Employing *in-situ* STEM on an all-solid-state electrochemical cell in a single experiment, we identify three distinct lithium orderings in $Li_xLaTe_3$, with $Li_3LaTe_3$ hosting a three-layer, tetragonal Li in the vdW gap. This is a new crystal structure for lithium which is body centered cubic in bulk. Such superdense lithium has not been observed in other layered materials via intercalation, except in free-standing bilayer graphene (*21*). Even in the bilayer graphene case, the exact lithium ordering was not known. Our discovery is made possible by multimodal STEM where we acquire several types of data in a single experiment, including HAADF-STEM images, EELS spectrum images, and 4D-STEM datasets during lithiation. This is akin to multiple synchrotron experiments on separate beamlines, but with the added advantage of the high spatial resolution of STEM and direct tracking of lithium with EELS. Thus, our methodology establishes a path for streamlined discovery and observation of intercalation-driven phase transitions.

**Acknowledgements**
This work made use of the Cornell Center for Materials Research shared instrumentation facility, Helios FIB, and Thermo Fisher Spectra 300 STEMs whose acquisitions were partly supported by the National Science Foundation (NSF) DMR-2039380 and the Platform for the Accelerated Realization, Analysis, and Discovery of Interface Materials (PARADIM) Award No. DMR-2039380. This work was performed in part at the Cornell NanoScale Facility, a member of the NSF National Nanotechnology Coordinated Infrastructure (NNCI), Grant NNCI-2025233. N.L.W. acknowledges additional support from the NSF GRFP under Award No. DGE- 2139899. *In-situ* STEM experiments were supported by DOE DE-SC0023905. Work at Princeton was supported by the Air Force Office of Scientific Research under award number FA9550-24-1-0110.

**Author contributions**

‖ N.L.W. and S.D.F. contributed equally to this work. N.L.W. and J.J.C conceived the project. N.L.W, S.D.F., and S.L. carried out the multi-modal STEM experiments and analyzed the data. S.S., T.H., and S.D.F. performed additional STEM experiments and simulated electron diffraction patterns. S.X. and N.L.W. performed Raman measurements. D.N., J.K., N.A.B., and T.A.A. performed *Ab initio* calculations. M.F. and L.A.A. assembled and tested the coin-type cells. J.L., R.S., and L.M.S. grew the $LaTe_3$ crystals. N.L.W, S. L. and G.S. made devices. N.L.W, S.D.F, and J.J.C wrote the manuscript with input from all authors. All authors have given approval to the final version of the manuscript.